# IN WHAT CONDITION CAN THE 125 GEV HIGGS BOSON DECAY TO A PAIR OF ON-SHELL W-BOSONS?

**Rasmiyya E. Gasimova**[1,2,3], **Vali A. Huseynov**[2,3,4]

[1] Shamakhy Astrophysical Observatory National Academy of Sciences of Azerbaijan, Y. Mammadaliyev Settlement, Shamakhy District, AZ5618, Azerbaijan;
[2] Department of Theoretical Physics, Baku State University, Z. Khalilov 23, AZ 1148, Baku, Azerbaijan;
[3] Department of Physics, Qafqaz University, Baku-Sumgayit Road, 16 km., Khirdalan, Baku, AZ0101, Azerbaijan;
[4] Department of General and Theoretical Physics, Nakhchivan State University, University Campus, AZ 7012, Nakhchivan, Azerbaijan;
E-mail: gasimovar@yahoo.co.uk
E-mail: vgusseinov@yahoo.com

**ABSTRACT**

We have determined that the decay of the neutral boson at a mass around $125\ GeV$ into an on-shell $W^-W^+$-pair in a uniform magnetic field becomes, in principle, possible and the new decay channel of this boson in a magnetic field is allowed by the energy and total angular momentum conservation laws. The required magnetic field strength for observation of the measurable effect is $\sim 10^{23}\ G$ (or in Teslas $\sim 10^{19}\ T$). The existence of the other neutral boson with the spin $J = 0$ and with the other mass is not either excluded in the mass range below $2m_W \cong 160.77\ GeV$.

Recently a new neutral boson (NB) at a mass around $125\ GeV$ [1, 2] with properties compatible with the Standard Model Higgs boson was discovered in the ATLAS and CMS experiments. This boson is described with $J^{PC} = 0^{++}$ where $P$ is the parity, $C$ is the charge conjugation, $J$ is the spin. One of the decay channels of the discovered Higgs-like boson is $H \to WW^*$. One of these $W^\mp$-bosons is on-shell, the other one ($W^*$) is off-shell. According to the energy conservation law the decay of this NB into the on-shell $W^-W^+$-boson pair is impossible because of $m_H < 2m_W$, where $m_W \cong 80.385\ GeV$ [3] is the $W$-boson mass. Therefore, this NB decays into one on-shell $W$-boson and one off-shell $W^*$-boson. The following natural questions arise. In what condition can the $125\ GeV$ Higgs boson decay to a pair of on-shell $W$ -bosons? How is realistic and promising the decay of the NB at a mass around $125\ GeV$ into the two on-shell $W$-bosons in a magnetic field (MF)? Search for the answers to these questions determines the motivation for the presented investigation. We investigate the decay of the NB at a mass around $125\ GeV$ observed at the LHC into a pair of on-shell $W$-bosons in a MF. The main purpose of this work is to determine the condition in what the NB at a mass around $125\ GeV$ can decay to a pair of on-shell $W$-bosons?

The energy of a $W^{\mp}$-boson in a homogenous MF is given by the formula [4, 5]

$$E_{W^{\mp}}^2 = p_{\mp z}^2 + (2n_{\mp} + 1 - 2q_{\mp}s_{\mp z})eB + m_W^2, \qquad (1)$$

where $B = |\vec{B}|$ is the strength of a MF whose intensity vector $\vec{B}$ is directed along the $Oz$-axis, $p_{\mp z}$ ($s_{\mp z}$) are the third component of the momentum (spin) of a $W^{\mp}$-boson, $n_{\mp} = 0, 1, 2, ...$ enumerates the Landau energy levels of a $W^{\mp}$-boson, $q_{-} = -1$ ($q_{+} = +1$) is the sign of the electric charge of a $W^{-}(W^{+})$-boson. A $W^{\mp}$-boson has three polarization states: $|W^{\mp}(s_{\mp} = 1, s_{\mp z} = +1)\rangle = |1, +1\rangle$, $|W^{\mp}(s_{\mp} = 1, s_{\mp z} = 0)\rangle = |1, 0\rangle$, $|W^{\mp}(s_{\mp} = 1, s_{\mp z} = -1)\rangle = |1, -1\rangle$, where $s_{\mp}$ is the spin of a $W^{\mp}$-boson. Hereafter we will consider the case $n_{\mp} = 0$, $p_{\mp z} = 0$ and $q_{\mp}s_{\mp z} = +1$ that corresponds to the ground Landau level of the $W^{-}(W^{+})$-boson. When $s_{-z} = -1$ and $s_{+z} = +1$, we only have dealings with the states $|W^{-}(s_{-} = 1, s_{-z} = -1)\rangle = |1, -1\rangle$ and $|W^{+}(s_{+} = 1, s_{+z} = +1)\rangle = |1, +1\rangle$. In this case the $W^{\mp}$-boson energy satisfies the inequality $E_{W^{\mp}} = \sqrt{m_W^2 - eB} < m_W$ for an arbitrary $B$ taken from the range $0 < B < B_{0W}$ where $B_{0W} = m_W^2/e$.

One of the main decay modes of the NB at a mass around $125\ GeV$ is the $H \to WW^* \to l\nu l\nu$ channel [6]. According to the energy conservation law the decay of this NB into the on-shell $W^{-}W^{+}$-boson pair is impossible. Therefore, this NB decays into one on-shell $W$-boson and one off-shell $W^*$-boson. However, if we place this NB in a uniform MF, the MF will affect on the $W^{-}$- and $W^{+}$-bosons. If we consider the NB at a mass around $125\ GeV$ in the rest frame and take into account the relation $E_{W^{\mp}} = \sqrt{m_W^2 - eB} < m_W$ for the $W^{-}(W^{+})$-boson with $s_{-z} = -1$ ($s_{+z} = +1$) in the energy conservation law $m_H = E_H = E_{W^{-}} + E_{W^{+}}$, we can see that in a sufficiently strong MF with the strength $B_H$ the equality $m_H = 2\sqrt{m_W^2 - eB_H}$ is satisfied and the decay of the NB with the mass around $125\ GeV$ into the two on-shell $W^{\mp}$-bosons becomes, in principle, energetically possible. So, as a result of the decay of $125\ GeV$ NB in a MF we have the final diboson system $W^{-}W^{+}$ that consists of the on-shell $W^{-}$- and $W^{+}$-bosons situating in a MF. The quantum states of the system $W^{-}W^{+}$ can have the spin equal to $0$, $1$, $2$. Using the two polarization states $|W^{-}(s_{-} = 1, s_{-z} = -1)\rangle = |1, -1\rangle$ and $|W^{+}(s_{+} = 1, s_{+z} = +1)\rangle = |1, +1\rangle$ of $W^{\mp}$-bosons and the addition rule of spins we obtain three polarization states of the system $W^{-}W^{+}$: $|0, 0\rangle$, $|1, 0\rangle$, $|2, 0\rangle$. It should be noted that in case of the longitudinal polarization of the spin of the $W^{\mp}$-boson we have $s_{\mp z} = 0$ and from the formula (1) we derive $E =$ $\sqrt{m_W^2 + eB} > m_W$ for an arbitrary $B$ taken from the range $0 < B < B_{0W}$ and obtain for the mass of the decaying NB the relation $m = 2\sqrt{m_W^2 + eB} > 2m_W$ that contradicts to the condition $m_H \cong 125\ GeV < 2m_W$. So, in the mass range $0 < m < 2m_W$ there is no sense to investigate the case of the longitudinal polarization of the spin of the on-shell $W^-(W^+)$-boson in a MF. According to the energy conservation law the energy of the final $W^-W^+$-system $E_{W^-W^+}$ is to be in the range $0 < E_{W^-W^+} < 2m_W$. The polarization states $|W^-(s_- = 1, s_{-z} = -1)\rangle = |1,-1\rangle$ and $|W^+(s_+ = 1, s_{+z} = +1)\rangle = |1,+1\rangle$ that form the three possible final states $|0,0\rangle, |1,0\rangle, |2,0\rangle$ only satisfy the condition $0 < E_{W^-W^+} < 2m_W$. The energy conservation law is as $m = m_H = 2\sqrt{m_W^2 - eB_H}$ when $W^-(s_- = 1, s_{-z} = -1)$ and $W^+(s_+ = 1, s_{+z} = +1)$ are produced on the ground Landau level. Since we consider a NB in the rest frame, the mass of the decaying NB can not be zero: $m \neq 0$. Taking into account the condition $m \neq 0$ we obtain from the formula $m = m_H = 2\sqrt{m_W^2 - eB_H}$ that $B_H \neq B_{0W}$ and $B_H < B_{0W}$. So, the mass of the NB at a mass around $125\ GeV$ satisfies the condition $0 < m_H < 2m_W$ ($0 < m_H < 160.77\ GeV$) for an arbitrary $B$ taken from the range $0 < B_H < B_{0W}$.

We can determine the charge conjugation $C_{W^-W^+}$, the parity $P_{W^-W^+}$ and the total angular momentum $J$ of the $W^-W^+$-system by the formulae:

$$C_{W^-W^+} = (-1)^{L_{W^-W^+} + S_{W^-W^+}}, \qquad (2)$$

$$P_{W^-W^+} = (-1)^{L_{W^-W^+}} P_{W^-} P_{W^+} = (-1)^{L_{W^-W^+}}, \qquad (3)$$

$$J = L_{W^-W^+} + S_{W^-W^+}, L_{W^-W^+} + S_{W^-W^+} - 1, \ldots, |L_{W^-W^+} - S_{W^-W^+}| \qquad (4)$$

where $P_{W^-}(P_{W^+})$ is the intrinsic parity for the $W^-(W^+)$-boson, $L_{W^-W^+}$ ($S_{W^-W^+}$) is the orbital quantum number (the total spin) for the $W^-W^+$-system. We take into account that the NB at a mass around $125\ GeV$ also decays into the two photons. Therefore its spin $J$ can not be 1 according to the Landau-Yang theorem [7, 8] and the charge conjugation $C$ of this NB is $C = C_H = 1$. The decay $H \to W^-W^+$ is a weak process and $C$ is not conserved in this process. It means that if the charge conjugation of the initial neutral $H$-boson is $C_H = 1$ before the reaction $H \to W^-W^+$, the charge conjugation $C_{W^-W^+}$ might be $+1$ or $-1$ after the reaction, or it might also go to a state that is not a $C_{W^-W^+}$ eigenstate. Here we assume that $C_{W^-W^+}$ (and $P_{W^-W^+}$) is either $+1$ or $-1$ after the reaction. The following combinations of $C_{W^-W^+}$ and $P_{W^-W^+}$ for the $W^-W^+$-system are possible: A) $C_{W^-W^+} = +1, P_{W^-W^+} = +1$; B) $C_{W^-W^+} = +1, P_{W^-W^+} = -1$; C) $C_{W^-W^+} =$ $-1, P_{W^-W^+} = +1$; D) $C_{W^-W^+} = -1, P_{W^-W^+} = -1$. We have obtained that only in the cases A and B the NB with the spin $J = 0$ can exist in the mass range $0 \leq m_H \leq 2m_W$ and it can decay into the on-shell $W^-(s_- =1, s_{-z} = -1)$- and $W^+(s_+ = 1, s_{+z} = +1)$-bosons in a MF:

case A: if $C_{W^-W^+} = +1$ and $P_{W^-W^+} = +1$, $J = 0, 2$ ($S_{W^-W^+} = 0, 2; L_{W^-W^+} = 0$), (10)

case B: if $C_{W^-W^+} = +1$ and $P_{W^-W^+} = -1$, $J = 0, 2$ ($S_{W^-W^+} = 1; L_{W^-W^+} = 1$), (11)

We have obtained $J = 0, 2$ for the spin of the NB at a mass around $125\,GeV$ if $C_{W^-W^+} = +1$. One NB with the spin $J = 0$ has already been observed in the mass range $0 < m_H \leq 2m_W$ by the ATLAS and CMS Collaborations [1, 2]. However, the existence of the other NB with the spin $J = 0$ and with the other mass is not excluded in the mass range $0 < m_H \leq 2m_W$. The analysis of the above considered cases A, B, C and D show that the existence of the NB at a mass around $125\,GeV$ with the spin $J = 0$ is allowed in the cases A and B. When $W^-W^+$-pair are produced on the ground Landau level, the orbital quantum number $L_{W^-W^+}$ should be minimal. $L_{W^-W^+}$ is minimal only in the case A. So, the case A is more suitable for the particle with the spin $J = 0$. The MF strength required for the decay of the NB at a mass around $125\,GeV$ into the on-shell $W^-(s_- = 1, s_{-z} = -1)$- and $W^+(s_+ = 1, s_{+z} = +1)$-bosons is calculated by the formula $B_H = B_{0W}[1 - (m_H/2m_W)^2]$. Let us perform the simple numerical estimations for the strength of the MF required for the decay of the NB at a mass around $125\,GeV$ into the on-shell $W^-(s_- = 1, s_{-z} = -1)$- and $W^+(s_+ = 1, s_{+z} = +1)$-bosons in a MF. When $m \cong 125\,GeV$, the required MF strength is $\sim 10^{23}\,G$ (or in Teslas $\sim 10^{19}\,T$). The strongest (pulsed) MF ever obtained in a laboratory is $28\,MG$ [9] that is much less than $\sim 10^{22}\,G$. The maximum strength of the produced strong MF in noncentral heavy-ion collisions in the direction perpendicular to the reaction plane is estimated to be $\sim 10^{17}\,G$ at the RHIC and $\sim 10^{18}\,G$ at the LHC [10-16]. In lead-lead collisions at the LHC, the strength of the generated MF may reach $\sim 10^{20}\,G$ [11, 12]. We hope that in the future collider experiments, when the strength of the produced strong MF reaches the magnitude $\sim 10^{22 \div 23}\,G$, the decay of the NB at a mass around $125\,GeV$ into the on shell $W^-$-and $W^+$-bosons can be observed experimentally. The decay of the NB at a mass around $125\,GeV$ into the on shell $W^-$- and $W^+$-bosons, probably, could be realized in the early stages after the Big Bang when sufficiently strong MFs existed.

We have determined that the decay of the NB at a mass around 125 $GeV$ into an on-shell $W^-W^+$-pair in a uniform MF becomes, in principle, possible and the new decay channel of this boson in a MF is allowed by the energy and total angular momentum conservation laws. The required MF strength for observation of the measurable effect is $\sim 10^{23}G$ (or in Teslas $\sim 10^{19}T$). The existence of the other NB with the spin $J=0$ and with the other mass is not either excluded in the mass range below $2m_W \cong 160.77 GeV$. We hope that the possibility of the decay of the NB at a mass around 125 $GeV$ into a pair of on-shell $W$-bosons in a uniform MF will be of interest of cosmologists and particle astrophysicists and it will also attract the experimental physicists' attention in future collider experiments.

## ACKNOWLEDGEMENTS

R. G. and V. H. are very grateful to the Organizing Committee of the LHCP2014 Conference for the kind invitation to attend this conference.